\documentclass[twocolumn]{IEEEtran}
\normalsize
\usepackage[cmex10]{amsmath}
\usepackage{graphicx}
\usepackage{epstopdf}
\usepackage[noadjust]{cite}
\begin{document}
\title{Conditional Entropy based User Selection for Multiuser MIMO Systems}
\author{Gaurav~Gupta and A.K.~Chaturvedi,~\IEEEmembership{Senior~Member,~IEEE}
\thanks{Gaurav Gupta and A.K. Chaturvedi are with the Department of Electrical Engineering, Indian Institute of Technology Kanpur, India (email: gauravgg@iitk.ac.in; akc@iitk.ac.in).}}
\markboth{IEEE Communications letters (Accepted)}{Accepted}
\maketitle 
%
%---------------------------------------------------------------------------
%-------------------------------ABSTRACT------------------------------------
%---------------------------------------------------------------------------
%
\begin{abstract}

We consider the problem of user subset selection for maximizing the sum rate of downlink multi-user MIMO systems. The brute-force search for the optimal user set becomes impractical as the total number of users in a cell increase. We propose a user selection algorithm based on conditional differential entropy. We apply the proposed algorithm on Block diagonalization scheme. Simulation results show that the proposed conditional entropy based algorithm offers better alternatives than the existing user selection algorithms. Furthermore, in terms of sum rate, the solution obtained by the proposed algorithm turns out to be close to the optimal solution with significantly lower computational complexity than brute-force search.

\end{abstract}
\IEEEpeerreviewmaketitle
%
%----------------------------------------------------------------------
% SECTION I: Introduction
%----------------------------------------------------------------------
%
\section{Introduction}
\label{sec:intro}

\IEEEPARstart{I}n Multiuser MIMO (MU-MIMO) systems the base station broadcasts to multiple users simultaneously with different data for different users, which gives rise to inter-user interference. Given the complexity of optimal Dirty Paper Coding (DPC), several linear suboptimal techniques such as Zero-Forcing Beamforming (ZFBF), Block Diagonalization (BD) \cite{spencer, choi} etc. have been proposed to cancel inter-user interference. ZFBF uses weight vectors which are chosen to cancel the interference among user streams. On the other hand, BD exploits the null space of channel space of the other users' using Singular Value Decomposition (SVD). The data meant to be transmitted to a particular user is multiplied by a precoding matrix which lies in the null space of channel spaces of other users being served simultaneously. Due to rank and nullity constraints the number of users which can be simultaneously supported are limited by the number of transmit and receive antennas. This leads to the problem of selecting the subset of users which can maximize the sum rate, which we refer to as the optimal subset of users. In a system where the number of users is large, the brute-force determination of the optimal subset of users is prohibitive because of the large number of possible subsets and high computational complexity of SVD. To reduce the computation load many suboptimal algorithms have been proposed \cite{shen,zhang,lee}.

The authors of \cite{shen} proposed two suboptimal algorithms: c-algorithm and n-algorithm. At each step the c-algorithm selects the user which maximizes the sum rate while the n-algorithm selects the user which maximizes the channel frobenius norm. Their performance is close to optimal but their computational complexity is large as c-algorithm involves large number of SVD computations while n-algorithm involves heavy Gram-Schmidt Orthogonalization (GSO) computations. The authors of \cite{lee} proposed an algorithm based on chordal distance which is a measure of orthogonality between channel spaces. 

In this paper, we propose a conditional entropy based user selection algorithm. The algorithm uses sum conditional differential entropy as a measure to select users iteratively until the maximum number of simultaneously supportable users are selected. The rest of this paper is organized as follows. Section \ref{sec:sys model} introduces the system model and Section \ref{sec:condE} discusses the application of conditional differential entropy in a MU-MIMO setting. The proposed algorithm is described in Section \ref{sec:sch_algo}. Section \ref{sec:sim_result} presents the simulation results. Finally the conclusions are given in Section \ref{sec:concl}. 
%
%---------------------------------------------------------------------------
% Section II: System Model
%---------------------------------------------------------------------------
%
\section{System Model}
\label{sec:sys model}

In the considered MU-MIMO system the data stream after precoding is sent to M transmit antennas, resulting in a $M\times 1$ transmit vector. The channel is assumed to be slowly flat-fading. It is assumed that there is perfect Channel State Information at the Receiver (CSIR), and BS knows the channels of all the users perfectly i.e. there is Channel State Information at the Transmitter (CSIT). We assume $K_{T}$ users each with $N$ receive antennas. Thus

\begin{equation}
{\bf y}_{k} = {\bf H}_{k}{\bf x} + {\bf n}_{k},\quad k = 1,.....,K_{T}
\label{eqn:cmp_MU_MIMO}
\end{equation}

\noindent where ${\bf H}_k$  $(k=1,...,K_{T})$ is the $N\times M$ channel matrix for the $k\text{th}$ user, the entries of which are independently and identically distributed (i.i.d.) circular symmetric complex Gaussian random variables with zero mean and unit variance. Further, ${\bf n}_{k}$ is the $N\times 1$ complex Additive White Gaussian Noise (AWGN) vector with zero mean and unit variance i.i.d. entries, and ${\bf y}_k$ is the $N\times 1$ vector received by the $k\text{th}$ user. The transmitted vector ${\bf x}$ of size $M\times 1$, is given by

\begin{equation}
{\bf x} = \sum\limits_{i=1}^K {\bf T}_{i}{\bf s}_{i}
\label{eqn:trans_sig}
\end{equation}

\noindent where $K$ is the number of simultaneous users served by the BS, ${\bf s}_{i}$ is the $L\times 1$ data vector for the $i\text{th}$ user, preprocessed with $M\times L$ precoding matrix ${\bf T}_{i}$. The received signal for the $k\text{th}$ user in (\ref{eqn:cmp_MU_MIMO}) can be split into desired signal, interference from other users and AWGN originating at receiver, which is given by

\begin{equation}
{\bf y}_{k} = {\bf H}_{k}{\bf T}_{k}{\bf s}_{k} + \sum\limits_{i=1,i\neq k}^K {\bf H}_{k}{\bf T}_{i}{\bf s}_{i} + {\bf n}_{k}  
\label{eqn:ela_MU_MIMO}
\end{equation}

The problem of optimal user set selection on the basis of maximization of sum rate can be written as

\begin{equation}
\mathcal{R}_{opt} = \max\limits_{\mathcal{S} \subset \Gamma, |\mathcal{S}| \le K} \mathcal{R}\left(\mathcal{S}\right)
\label{eqn:opt_cap}
\end{equation}

\noindent where $\Gamma = \left\{1,...,K_{T}\right\}$, $\mathcal{R\left(\mathcal{S}\right)}$ is the sum rate of user set $\mathcal{S}$, $|\mathcal{S}|$ denotes cardinality of $\mathcal{S}$, $K$ is the maximum number of simultaneously supportable users by the considered MU-MIMO scheme and $\mathcal{R}_{opt}$ is the maximum possible sum rate. From (\ref{eqn:opt_cap}), we can see that the optimal scheduling algorithm selects a subset over all possible subsets of users subject to a cardinality constraint. 

The only known way to obtain the optimal solution is by performing brute-force search over all possible user subsets.
%
%---------------------------------------------------------------------------
% Section III: Sum Conditional Differential Entropy
%---------------------------------------------------------------------------
%
\section{Sum Conditional differential Entropy}
\label{sec:condE}

In this section we derive equations which will help in formulating the conditional entropy based algorithm later.

Let us consider a $n$ user MU-MIMO system

\begin{equation}
{\bf y}_{k} = {\bf H}_{k}{\bf x} + {\bf n}_{k}, \quad k = 1,...,n
\label{eqn:SU_MIMO}
\end{equation}

\noindent for which the information rate of the $k$th user $\mathcal{I}\left({\bf y}_{k}; {\bf x}\right)$ will be maximum when the differential entropy $\mathcal{H}\left({\bf y}_{k}\right)$ is maximum. With $E[{\bf x}{\bf x}^H] = {\bf Q}$ and power constraint $E[{\bf x}^H{\bf x}] \le P$, the distribution which maximizes $\mathcal{H}\left({\bf y}_{k}\right)$ is circular symmetric complex Gaussian \cite{telatar} and the differential entropy is given by

\begin{equation}
\mathcal{H}({\bf y}_{k}) = \text{log}_{2} \text{det}\left(\pi e\left({\bf H}_{k}{\bf Q}{\bf H}_{k}^H + {\bf I}_{N}\right)\right)
\label{eqn:MU_Ent}
\end{equation}

Now consider $\tilde{{\bf y}} = [{\bf y}_{1}^H,{\bf y}_{2}^H,\cdots,{\bf y}_{n}^H]^H$, $\tilde{{\bf y}}$ will also be a circular symmetric complex Gaussian random variable with zero mean and $E[\tilde{{\bf y}}\tilde{{\bf y}}^H] = {\bf \Sigma}$, such that ${\bf \Sigma} = \left[{\bf \Sigma}_{ij}\right]$ where

\begin{equation}
{\bf \Sigma}_{ij} = E[{\bf y}_{i}{\bf y}_{j}^H] = 
	\begin{cases}
	{\bf H}_{i}{\bf Q}{\bf H}_{i}^{H} + {\bf I}_{N} & \text{if~} i = j \\
	{\bf H}_{i}{\bf Q}{\bf H}_{j}^H & \text{if~} i \neq j
	\end{cases}
\label{eqn:sigma_def}
\end{equation}

The joint differential entropy of ${\bf y}_{k}$'s will be $\mathcal{H}(\tilde{{\bf y}}) = \text{log}_{2}\text{det}\left(\pi e{\bf \Sigma}\right)$ and can be written as

\begin{eqnarray}
\mathcal{H}(\tilde{{\bf y}})
 	&=&\text{log}_{2}\left((\pi e)^{nN}\times\right.\nonumber\\
 	 &&\left.\text{det}\left({\bf I}_{M} + {\bf Q}{\bf H}_{1}^H{\bf H}_{1} + \cdots +{\bf Q}{\bf H}_{n}^H{\bf H}_{n}\right)\right)
\label{eqn:sum_Ent}
\end{eqnarray}

\noindent where (\ref{eqn:sum_Ent}) has been written using matrix determinant identity

\begin{equation}
\text{det}\left({\bf I}_{M} + {\bf AB}\right) = \text{det}\left({\bf I}_{N} + {\bf BA}\right)
\label{eqn:mat_iden}
\end{equation}

\noindent where ${\bf A}$ and ${\bf B}$ are $M\times N$ and $N\times M$ matrices, respectively. The conditional differential entropy of ${\bf y}_{k}$ is given by

\begin{equation}
\mathcal{H}\left({\bf y}_{k} | \tilde{{\bf y}}_{k}\right) = \mathcal{H}\left(\tilde{{\bf y}}\right) - 
\mathcal{H}\left( \tilde{{\bf y}}_{k}\right)
\label{eqn:cond_Ent}
\end{equation}

\noindent where $\tilde{{\bf y}}_{k} = [{\bf y}_{1}^H,\cdots,{\bf y}_{k-1}^H,{\bf y}_{k+1}^H,\cdots,{\bf y}_{n}^H]^H$. Sum conditional differential entropy of $n$ random variables is defined as the sum of conditional differential entropy of each random variable with the other $n-1$ random variables. From (\ref{eqn:cond_Ent}) we can now write the sum conditional differential entropy of the users in $\mathcal{S} = \{1,2,...,n\}$ as

\begin{equation}
\mathcal{H}_{SC}\left(\mathcal{S}\right) = \sum\limits_{k \in \mathcal{S}} \mathcal{H}\left({\bf y}_{k} | \tilde{{\bf y}}_{k}\right)
\label{eqn:sum_cond_Ent}
\end{equation}
%
%---------------------------------------------------------------------------
% Section IV: Conditional Entropy based Scheduling Algorithm
%---------------------------------------------------------------------------
%
\section{Conditional Entropy based User Selection Algorithm}
\label{sec:sch_algo}

Precoding schemes for interference cancellation, for e.g. BD \cite{shen}, remove the common subspace between the channels of the selected users and hence entropy of the $k$th user's signal with effective channel ${\bf H}_{k}{\bf T}_{k}$ reduces. Therefore, for sum rate maximization we should attempt to select the users with not only maximum differential entropy but also with minimum common subspace. We know that as the channels' space tend to be orthogonal lesser is the subspace common to them. Hence we will bring in consideration of orthogonality.

A user selection algorithm using capacity upperbound as selection metric was
proposed in \cite{zhang}. It can be seen from (\ref{eqn:sum_Ent}) that capacity upperbound is identical to the joint differential entropy of selected users $\mathcal{S}$ and new user $t$. Thus, the formulation in \cite{zhang} seeks to maximize only the joint differential entropy and does not take orthogonality into account. Hence there is a possibility of sum rate improvement if we bring in consideration of orthogonality. We will show that the mutual information can serve this purpose.

Let us consider a MU-MIMO system (\ref{eqn:SU_MIMO}) with $n=2$. Then $\mathcal{I}\left({\bf y}_{1};{\bf y}_{2}\right)$ can be written as

\begin{IEEEeqnarray}{rCl}
\mathcal{I}\left({\bf y}_{1};{\bf y}_{2}\right) &=& \mathcal{H}\left({\bf y}_{1}\right) + \mathcal{H}\left({\bf y}_{2}\right) - \mathcal{H}\left({\bf y}_{1};{\bf y}_{2}\right)\nonumber\\
&=& \text{log}_{2}\text{det}\left(\vphantom{\sum}{\bf I} + {\bf Q}{\bf H}_{1}^H{\bf H}_{1}{\bf Q}{\bf H}_{2}^H{\bf H}_{2}\times\right.\nonumber\\
&&\left. \left({\bf I}_{M} + {\bf Q}{\bf H}_{1}^H{\bf H}_{1} + {\bf Q}{\bf H}_{2}^H{\bf H}_{2}\right)^{-1}\right)
\label{eqn:Inf_two_user}
\end{IEEEeqnarray}

\noindent where $\mathcal{H}\left({\bf y}_{1};{\bf y}_{2}\right)$ is the joint differential entropy of ${\bf y}_{1}$ and ${\bf y}_{2}$ written using (\ref{eqn:sum_Ent}) and the optimal value of ${\bf Q}$ is determined by BC-capacity region \cite{vishwa}. However, in order to bring in the consideration of orthogonality we will substitute ${\bf Q} = \frac{P}{M}{\bf I}$, so that ${\bf H}_{1}{\bf H}_{2}^H$ can appear in (\ref{eqn:Inf_two_user}). Hence

\begin{multline}
\mathcal{I}\left({\bf y}_{1};{\bf y}_{2}\right) = \text{log}_{2}\text{det}\left(\vphantom{\sum}{\bf I} + \left(P/M\right)^2{\bf H}_{1}^H{\bf H}_{1}{\bf H}_{2}^H{\bf H}_{2}\times\right.\\
\left. \left({\bf I}_{M} + P/M{\bf H}_{1}^H{\bf H}_{1} + P/M{\bf H}_{2}^H{\bf H}_{2}\right)^{-1}\right)
\label{eqn:Mut_Inf_mod}
\end{multline}

\noindent and $\mathcal{I}\left({\bf y}_{1};{\bf y}_{2}\right) = 0$ whenever the row spaces of ${\bf H}_{1}$ and ${\bf H}_{2}$ will be orthogonal i.e. ${\bf H}_{1}{\bf H}_{2}^H = 0$. In other words, mutual information between orthogonal users is zero. Therefore, lesser the mutual information, closer to orthogonality will be the users' channel. Now, to select two users from $\Gamma$, $s_{1}$ will be the user with maximum differential entropy. For selecting user $s_{2}$ from $t \in \Gamma - \{s_{1}\}$ we have to maximize $\mathcal{H}\left({\bf y}_{t};{\bf y}_{s_{1}}\right)$ and minimize $\mathcal{I}\left({\bf y}_{t};{\bf y}_{s_{1}}\right)$, which can be performed if we maximize $\mathcal{H}\left({\bf y}_{t};{\bf y}_{s_{1}}\right) - \mathcal{I}\left({\bf y}_{t};{\bf y}_{s_{1}}\right)$. Therefore

\begin{IEEEeqnarray}{lCl}
s_{2} &=& \text{arg} \max\limits_{t\in \Gamma-\{s_{1}\}}\left\{\mathcal{H}\left({\bf y}_{t};{\bf y}_{s_{1}}\right) - \mathcal{I}\left({\bf y}_{t};{\bf y}_{s_{1}}\right)\right\}\IEEEnonumber\\
&=& \text{arg} \max\limits_{t\in \Gamma-\{s_{1}\}}\left\{\mathcal{H}\left({\bf y}_{t}|{\bf y}_{s_{1}}\right) + \mathcal{H}\left({\bf y}_{s_{1}}\right) -\right. \IEEEnonumber\\ 
&&\left.\hspace*{80pt} \left(\mathcal{H}\left({\bf y}_{s_{1}}\right) - \mathcal{H}\left({\bf y}_{s_{1}}|{\bf y}_{t}\right)\right)\right\} \IEEEnonumber\\
&=& \text{arg} \max\limits_{t\in \Gamma-\{s_{1}\}}\left\{\mathcal{H}\left({\bf y}_{t}|{\bf y}_{s_{1}}\right) + \mathcal{H}\left({\bf y}_{s_{1}}|{\bf y}_{t}\right)\right\}
\label{eqn:arg_max_s2}
\end{IEEEeqnarray}

From (\ref{eqn:arg_max_s2}) we can see that $s_{2}$ is indeed the user with maximum sum conditional entropy (\ref{eqn:sum_cond_Ent}) given $s_{1}$. 
Thus sum conditional entropy implicitly includes orthogonality constraint, hence can be expected to give better performance than upperbound metric.
   
Using the above formulation, we will generalize the algorithm to the selection of more than two users. Let $\mathcal{S} = \left\{s_{1},...,s_{k}\right\}$ be the selected users after $k\text{th}$ user selection step. At $(k+1)\text{th}$ step, $s_{k+1}$ will be the user $t \notin \mathcal{S}$ which maximizes sum conditional differential entropy of the users in $\mathcal{S}$ and the user $t$ i.e. $\mathcal{H}_{SC}(\mathcal{S} + \{t\})$ in (\ref{eqn:sum_cond_Ent}). Therefore

\begin{align}
&s_{k+1} = \text{arg}\max\limits_{t \in \Gamma - \mathcal{S}}\mathcal{H}_{SC}(\mathcal{S} + \{t\}) \nonumber\\
	&\qquad= \text{arg}\max\limits_{t \in \Gamma - \mathcal{S}}\text{log}_{2} \nonumber\\
	&\text{det}\left({\bf I}_{N} + {\bf H}_{t}\left(
	\frac{M}{P}{\bf I}_{M} + {\bf H}(\mathcal{S})^H{\bf H}(\mathcal{S})\right)^{-1}
	{\bf H}_{t}^H\right) {+} \nonumber\\
	&\sum\limits_{i=1}^k 
	\text{log}_{2}
	\text{det}\left({\bf I}_{N} + {\bf H}_{s_{i}}\left(
	\frac{M}{P}{\bf I}_{M} + {\bf H}(\mathcal{S}_{i})^H{\bf H}(\mathcal{S}_{i})\right)^{-1}
	{\bf H}_{s_{i}}^H\right)
\label{eqn:sum_cEnt_mat}
\end{align}

\noindent where  $\mathcal{S}_{i} = \mathcal{S} + \{t\} -  \left\{s_{i}\right\}$. The $kN\times M$ channel matrices ${\bf H}\left(\mathcal{S}\right)$ and ${\bf H}\left(\mathcal{S}_{i}\right)$ are constructed by vertically aligning the channel matrices of the users in $\mathcal{S}$ and $\mathcal{S}_{i}$, respectively. The expression for $\mathcal{H}_{SC}(\mathcal{S} + \{t\})$ in (\ref{eqn:sum_cEnt_mat}) is written using (\ref{eqn:sum_Ent}), (\ref{eqn:mat_iden}) and (\ref{eqn:cond_Ent}) after dropping the constant term involving $(\pi e)$.

For this a series of matrix inversions are to be computed at each user selection step. This can be done through matrix inversion lemma or woodbury formula \cite{ben}. With ${\bf A}$ an $M\times N$ positive definite matrix and ${\bf B}$ an $N\times M$ matrix,

\begin{equation}
\left({\bf A} + {\bf B}^H{\bf B}\right)^{-1} = {\bf A}^{-1} - {\bf A}^{-1}{\bf B}^H\left({\bf I}_{N} + {\bf B}{\bf A}^{-1}{\bf B}^H\right)^{-1}{\bf B}{\bf A}^{-1}
\label{eqn:mat_inv_iden}
\end{equation}

Now we will derive recursion for the first term \cite{zhang} of (\ref{eqn:sum_cEnt_mat}) and the same arguments will be applied for the terms inside the summation in the same equation. Let us define ${\bf \Omega}_{k}$ by

\begin{equation}
{\bf \Omega}_{k} = \left(
	\frac{M}{P}{\bf I}_{M} + {\bf H}(\mathcal{S})^H{\bf H}(\mathcal{S})\right)^{-1}
\label{eqn:omega_for}
\end{equation}

We can write the effective channel at the $(k+1)\text{th}$ step as 

\begin{equation}
{\bf H}_{\text{eff}} = 
	\begin{bmatrix}
	{\bf H}(\mathcal{S})^H & {\bf H}_{t}^H\\
	\end{bmatrix}^H
\label{eqn:eff_channel}
\end{equation}

We now update (\ref{eqn:omega_for}) by replacing ${\bf H}(\mathcal{S})$ with ${\bf H}_{\text{eff}}$ from (\ref{eqn:eff_channel}) to obtain ${\bf \Omega}_{k+1}$ as

\begin{equation}
{\bf \Omega}_{k+1} = \left({\bf \Omega}_{k}^{-1} + {\bf H}_{t}^H{\bf H}_{t}\right)^{-1}
\label{eqn:omega_ite}
\end{equation}

On substituting ${\bf A}$ for ${\bf \Omega}_{k}^{-1}$ and ${\bf B}$ for ${\bf H}_{t}$ in (\ref{eqn:mat_inv_iden}), the recursion is given by

\begin{equation}
{\bf \Omega}_{k+1} = {\bf \Omega}_{k} - {\bf \Omega}_{k}{\bf H}_{t}^H\left({\bf I}_{N} + {\bf H}_{t}{\bf \Omega}_{k}{\bf H}_{t}^H\right)^{-1}{\bf H}_{t}{\bf \Omega}_{k}
\label{eqn:omega_algo}
\end{equation}

In Step $1$, the algorithm is initialized while in Step $2$, the algorithm first selects the user with maximum differential entropy and then successively selects the user which maximizes the sum conditional entropy till the maximum number of simultaneously supportable users limit $K$ is reached. The proposed algorithm is general in nature and is applicable to any scheme for which $\mathcal{R}\left(\mathcal{S}\right)$ can be calculated. Thus, when applied to different schemes, only the step involving $\mathcal{R}\left(\mathcal{S}_{temp}\right)$ will be different. In this paper $\mathcal{R}\left(\mathcal{S}_{temp}\right)$ is calculated using \cite{shen}. 

%
%---------------------------------------------------------------------------
%---------------------------PROPOSED ALGORITHM------------------------------
%---------------------------------------------------------------------------
%
\vspace{-3pt}
\hspace*{-10pt}\hrulefill \\
1) Initialization, $\Gamma = \left\{1,2,...,K_{T}\right\}, \mathcal{S} = \emptyset$. Let  ${\bf \Omega} = \frac{P}{M}{\bf I}_{M}$, ${\bf \Omega}_{n} = {\bf \Omega}$. Let $\mathcal{R}_{use} = 0;$ \\ [1ex]
2) for $i = 1: K$ \\ 
\hspace*{15pt} for $j = 1 : i - 1$ \\
\hspace*{25pt}For each $k \in \Gamma$, compute\\ 
\hspace*{25pt}${\bf \Omega}_{s_{j},k} = {\bf \Omega}_{s_{j}} - {\bf \Omega}_{s_{j}}{\bf H}_{k}^H\left({\bf I}_{N} + {\bf H}_{k}{\bf \Omega}_{s_{j}}{\bf H}_{k}^H\right)^{-1}{\bf H}_{k}{\bf \Omega}_{s_{j}}$; \\ 
\hspace*{15pt} end-for \\
\hspace*{15pt} $p = \text{arg}\max\limits_{k \in \Gamma}\{ \text{log}_{2}\text{det}\left({\bf I}_{N} + {\bf H}_{k}{\bf \Omega}_{n}{\bf H}_{k}^H\right) + $\\
\hspace*{15pt}\qquad $ \sum\limits_{s \in \mathcal{S}} 						 \text{log}_{2}\text{det}\left( {\bf I}_{N} + {\bf H}_{s}{\bf 					 \Omega}_{s,k}{\bf H}_{s}^H\right) \};$\\
\hspace*{10pt}$\quad \mathcal{R} = \mathcal{R} \left(\mathcal{S}_{temp} \right) ; \mathcal{S}_{temp} = \mathcal{S} + \{p\};$ \\
\hspace*{15pt} if $\mathcal{R}< \mathcal{R}_{use}$ \\ 
\hspace*{20pt} break; \\ 
\hspace*{15pt} else \\
\hspace*{20pt} $\mathcal{S} = \mathcal{S}_{temp} ; \Gamma = \Gamma - \{p\} ; \mathcal{R}_{use} = \mathcal{R};$ \\ 		 
\hspace*{15pt} end-if \\
\hspace*{15pt} $ {\bf \Omega}_{s_{i}} = {\bf \Omega}_{n}$ ; \\
\hspace*{15pt} $ {\bf \Omega}_{n} = {\bf \Omega}_{n} - {\bf \Omega}_{n}{\bf H}_{p}^H\left({\bf I}_{N} + {\bf H}_{p}{\bf \Omega}_{n}{\bf H}_{p}^H\right)^{-1}{\bf H}_{p}{\bf \Omega}_{n};$ \\ 
\hspace*{5pt}end-for

\vspace*{-9pt}
\hspace*{-10pt}\hrulefill
%
%---------------------------------------------------------------------------
%

The matrix inversion lemma (\ref{eqn:mat_inv_iden}) is known to be numerically unstable when a large number of recursions are performed, for e.g. in adaptive filtering. However in our algorithm the recursions to update the ${\bf \Omega}_{k}$ are being done for a maximum of $K$ times. Since $K$ is small and does not increase with $K_{T}$, numerical stability of the matrix inversion lemma is not an issue.

The flop count of the algorithm is as follows: For computing the positive definite matrix ${\bf I}_{N} + {\bf H}_{k}{\bf \Omega}{\bf H}_{k}^{H}$, its determinant and inverse using Cholesky decomposition the flops required are $8M^2N + 8MN^2 + N$, $\frac{4}{3}N^3 - \frac{3}{2}N^2 + \frac{13}{6}N$ \cite{zhang} and $4N^3 -\frac{1}{2}N^2 - \frac{3}{2}N$ \cite{hunger} respectively. Hence to update ${\bf \Omega}$ in (\ref{eqn:omega_algo}), flops required are $\psi_{\Omega} = 32M^2N + 16MN^2 + 2M^2 + N + 4N^3 -\frac{1}{2}N^2 - \frac{3}{2}N$. Therefore, the total flops of the algorithm is

\begin{eqnarray}
\psi_{CE} &\approx & \sum\limits_{i=1}^K \left\{\psi_{\Omega}\times (i-1) + [\tfrac{4}{3}N^3 - \tfrac{3}{2}N^2 + \tfrac{19}{6}N \right. \nonumber \\
&&\quad\left.{+}\: 8M^2N + 8MN^2]\times i \right\} \times \left(K_{T} - i + 1\right) \nonumber \\
&& {+}\: K\times \psi_{\Omega}\nonumber \\
&\approx & \mathcal{O}\left(K_{T}KM^3\right)
\label{eqn:compl_PA}
\end{eqnarray}
%
%---------------------------------------------------------------------------
% Section V: Simulation Results
%---------------------------------------------------------------------------
%
\section{Simulation Results}
\label{sec:sim_result}

In this section, we provide the sum rate and flop count results for the proposed conditional entropy based user selection algorithm when applied to BD scheme. We compare the sum rate achieved by the proposed algorithm with the optimal solution and the existing algorithms namely, c-algorithm, n-algorithm, upperbound based algorithm, chordal distance based algorithm and row-norm based algorithm \cite{tran}.

\begin{figure}
\centering
\includegraphics*[viewport=345 10 857 447, width = 3.2in, height = 2.7in]{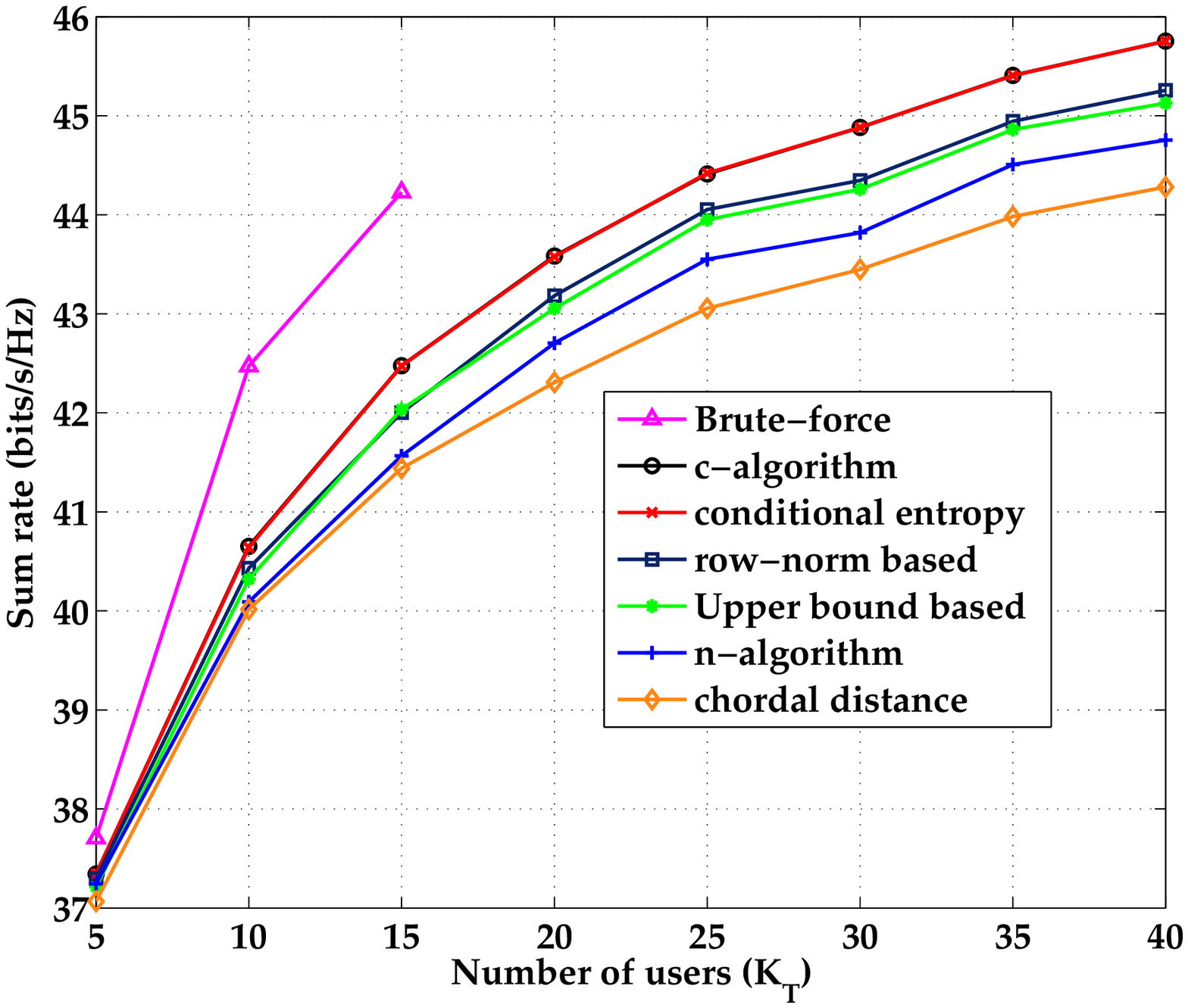}
\caption{Sum rate versus total number of users when $M = 8, N = 2, K = 4$ for $\text{SNR} = 20\text{~dB}$.}
\label{fig:per_20dB_8_2}
\end{figure}

\begin{figure}
\centering
\includegraphics*[viewport=330 10 857 447, width = 3.2in, height = 2.7in]{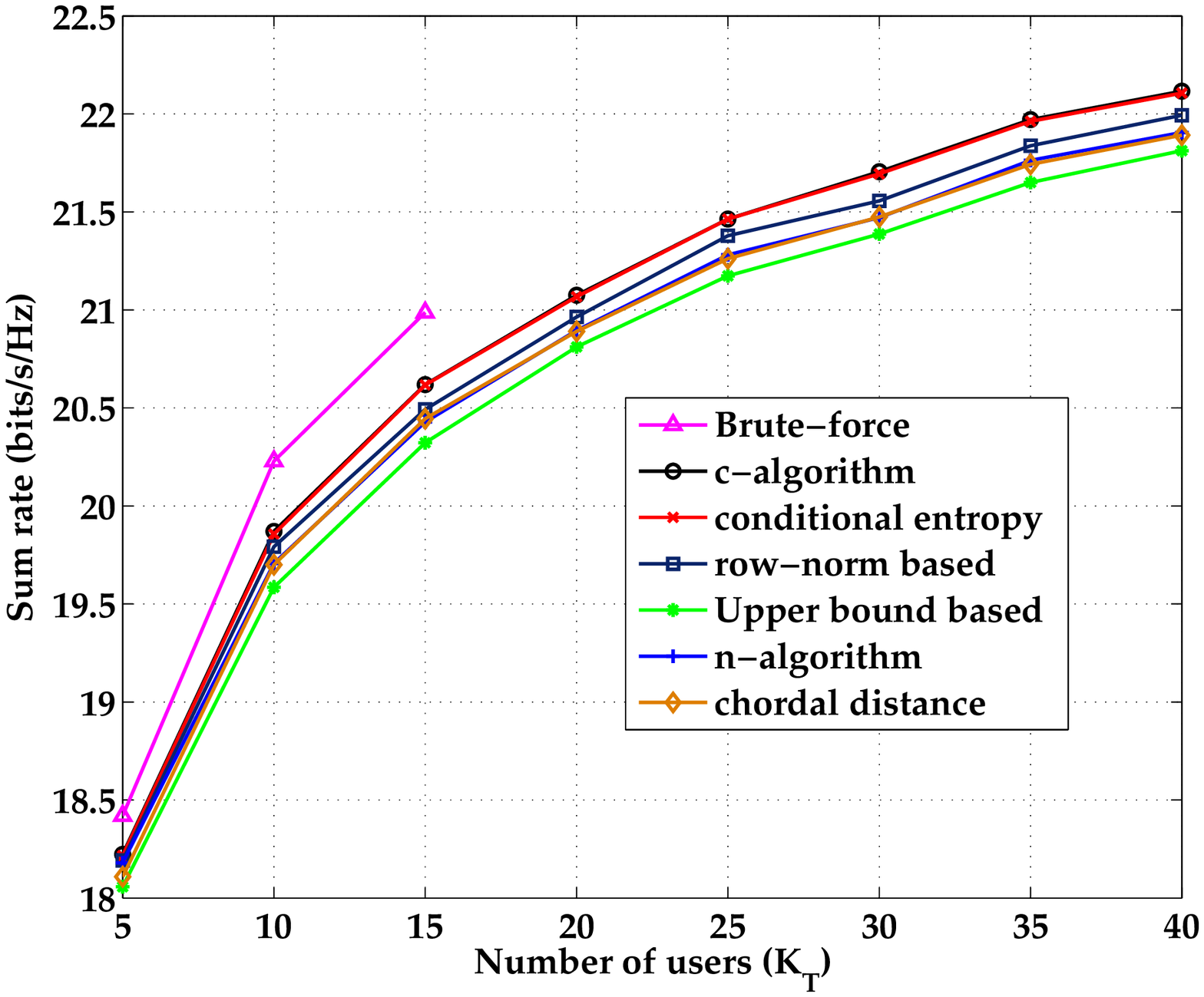}
\caption{Sum rate versus total number of users when $M = 8, N = 2, K = 4$ for $\text{SNR} = 10\text{~dB}$.}
\label{fig:per_10dB_8_2}
\end{figure}

In \figurename\,\ref{fig:per_20dB_8_2} and \figurename\,\ref{fig:per_10dB_8_2}, we compare the sum rate versus the total number of users $\left(K_{T}\right)$ for $(M, N) = (8, 2)$, i.e. $K = 4$, at $\text{SNR} = 20\text{~dB}$ and $10\text{~dB}$ respectively. We can see that the sum rate of conditional entropy based algorithm is strictly better than n-algorithm, row-norm based algorithm, upperbound based algorithm and chordal distance based algorithm. Moreover, we can observe that the plots of c-algorithm and conditional entropy based algorithm are overlapping\footnote{This is why only six curves are visible even though seven curves have been plotted.} and achieve approximately $95\%$ sum rate of the optimal solution.

\begin{figure}
\centering
\includegraphics*[viewport=360 7 836 425, width = 3.2in, height = 2.7in]{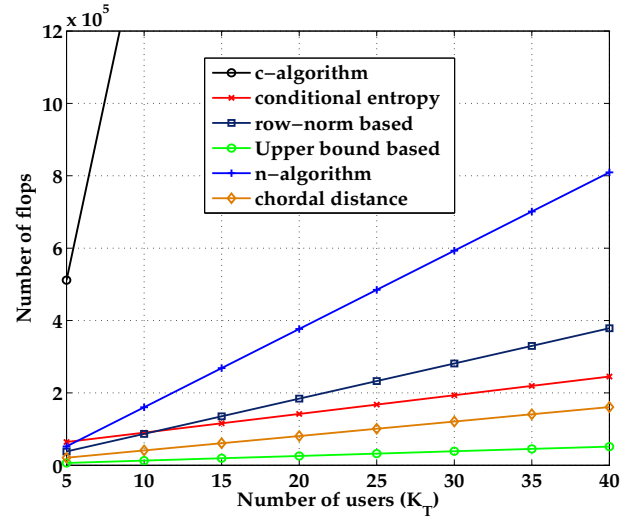}
\caption{Number of flops versus total number of users when $M = 8, N = 2, K = 4$}
\label{fig:flops_8_2}
\end{figure}

In \figurename\,\ref{fig:flops_8_2} we show the total number of flops versus $K_{T}$ of all these algorithms for $\left(M, N\right) = (8, 2)$. It can be observed that the c-algorithm has highest flop count. Further, it should be noted that even though the sum rate plots of the c-algorithm and the conditional entropy based algorithm overlap, flop count of the conditional entropy based algorithm is significantly lower. It may be noted that the chordal distance based algorithm and upperbound based algorithm have a lower flop count but it  comes at the cost of their lower sum rate as observed in \figurename\,\ref{fig:per_20dB_8_2} and \figurename\,\ref{fig:per_10dB_8_2}.
%
%---------------------------------------------------------------------------
% Section VII: Conclusion
%---------------------------------------------------------------------------
%
\section{Conclusion}
\label{sec:concl}

Although we have shown the conditional entropy based algorithm only for BD scheme, the algorithm is potentially applicable to any other MU-MIMO scheme like Successive Zero-forcing \cite{dabbagh}. The simulation results show that the proposed algorithm achieves higher sum rate and/or lower complexity than the existing algorithms. Also, the sum rate obtained by the proposed algorithm is close to that achieved by brute-force search based optimal algorithm, with significantly lower complexity.
%
%---------------------------------------------------------------------------
%--------------------------------REFERENCE----------------------------------
%---------------------------------------------------------------------------
%
\bibliographystyle{IEEEtran}
\bibliography{IEEEabrv,condE}
%
%---------------------------------------------------------------------------
%-----------------------------------END-------------------------------------
%---------------------------------------------------------------------------
%
\end{document}